\input{aipcheck}

\documentclass[
    ,final            
  ]
  {aipproc}

\layoutstyle{8x11double}

\begin{document}

\title{Search for VHE $\gamma$-ray emission in the vicinity of selected pulsars of the Northern Sky with VERITAS}

\classification{95.85.Pw,95.85.Nv,98.54.Cm}
\keywords      {Pulsar Wind Nebulae:individual(3C58,Geminga,J0631+1036,J1740+1000,B0656+14),gamma rays: observations, VERITAS}

\author{Ester Aliu$^{1}$ for the VERITAS Collaboration$^{2}$}{
  address={{\small $^1$Department of Physics and Astronomy and the Bartol Research Institute, University of Delaware, DE 19716, USA}}
  ,altaddress={$^2$ http://veritas.sao.arizona.edu} 
}

\begin{abstract}
It is generally believed that pulsars dissipate their rotational energy through powerful winds of relativistic particles. Confinement of these winds leads to the formation of luminous pulsar wind nebulae (PWNe) seen across the electromagnetic spectrum in synchrotron and inverse Compton emission. Recently, many new detections have been produced at the highest energies by Very High Energy (VHE) $\gamma$-ray observations identifying PWNe as among the most common sources of galactic VHE $\gamma$-ray emission. We report here on the preliminary results of a search for VHE $\gamma$-ray emission towards a selection of energetic and/or close pulsars in the Northern hemisphere in the first year of operations of the full VERITAS array.
\end{abstract}

\maketitle


\section{Introduction}
The detection of VHE $\gamma$-rays from the Crab Nebula\cite{Weekes89} together with the detection of at least 8 other PWNe in the HESS VHE $\gamma$-ray survey of the southern Galactic plane\cite{Aharonian05a}, has established PWNe as strong VHE $\gamma$-ray emitters. 
\par
Our basic understanding of a PWN consists of a magnetized wind carrying relativistic charged particles which emerges from a pulsar. The particles are injected into the pulsar's immediate surroundings, which are generally the interior of a Supernova Remnant (SNR), except when the age of the remnant is large. The confinement of this wind in the outer regions of the PWNe is thought to produce a wind termination shock in which the particles of the wind are accelerated and have their pitch angles randomized. As a result, bright synchrotron emission is seen in the radio through the low energy $\gamma$-ray range, as well as VHE $\gamma$-ray emission via the inverse Compton scattering of ultrarelativistic electrons on ubiquitous photon fields, such as the Cosmic Microwave Background (CMBR) and possibly infrared radiation from galactic dust particles. 
\par
The details of the wind nebulae structure and luminosity depend upon a number of factors, including the birth properties of the pulsar, the composition of the pulsar wind, the spindown power history and space velocity of the pulsar, as well as the density of the surrounding medium. Therefore, it is important to accumulate as much data as possible, at all wavelengths, in order to constrain models of these sources and produce a better understanding of which, how and when particles are accelerated and of the emission mechanisms taking place. 
\par
Young pulsars with high spindown power, similar to that powering the Crab Nebula, are considered among the best candidates to drive detectable VHE $\gamma$-ray nebulae. A significant correlation has now been suggested between VHE $\gamma$-ray emission above 100 GeV and pulsar energy output in terms of $\dot{E}/d^2$ \cite{Carrigan07} from observations at Southern latitudes (from -60$^{\circ}$ to 30$^{\circ}$ in Galactic longitude and $\pm$2 deg in Galactic latitude). Of the total of 435 pulsars in that region, 30 are found with significant VHE $\gamma$-ray emission. The probability that the detection of VHE sources coincident with 9 or more of the total of 23 pulsars above $\dot{E}/d^2>10^{34}$ ergs/s/kpc$^2$ results from a statistical fluctuation is $\sim3.4\times10^{-4}$. For detection of 5 or more of the total of 7 pulsars above $10^{35}$ ergs/s/kpc$^2$, the chance probability is $\sim4.2\times10^{-4}$. Motivated by this positive correlation, that argues in favour of a pulsar-related origin of the $\gamma$-ray signal, a search for VHE $\gamma$-ray counterparts at Northern latitudes has been started with VERITAS. Here we present some initial results of this search; further observations and analysis improvements are ongoing. 

\subsection*{Selection of the objects}
The criteria used for the selection of objects in this study is pulsars with $\dot{E}/d^2>1\times10^{35}$ ergs/s/kpc$^2$. In this range, the aforementioned population studies of the observations at Southern latitudes show that 70\% of the pulsars correlate with sources of VHE $\gamma$-rays. Additionally none of these ojects has to lie within the Galactic plane as this is being surveyed with VERITAS. Using these two criteria a handful of objects is available. The observations presented here consists of 5 objects which are here listed:
\begin{description}
\item[J0205+6449:] Pulsar with high spindown power $\dot{E}=2.5\times10^{37}$ erg/s, d = 3.2 kpc.The PWN is better known as 3C58 and has been detected in radio, infrared and X-rays with a size of about 9'$\times$5'. 
\item[J0631+1036:] pulsar with spindown power $\dot{E}=1.7\times10^{35}$ erg/s and d = 6.55 kpc. 
\item[J0633+1746:] bow shock PWN, pulsar with spindown power $\dot{E}=3.25\times10^{34}$ and d = 0.16 kpc, possible association with an extended MILAGRO hot spot C3.
\item[B0656+14:] pulsar with spindown power $\dot{E}=3.8\times10^{34}$ erg/s and at a short distance from us d = 0.29 kpc, very close in projection to the center of the Monogem ring
\item[J1740+1000:]pulsar with spindown power $\dot{E}=2.3\times10^{35}$ erg/s and as close as d = 1.36 kpc.
\end{description}
PSR J0633+1746 corresponds to Geminga and is here only mentioned for completeness as it satisfies the same selection criteria as the other sources. A more detailed study on Geminga can be found also in these proceedings~\cite{Kieda08}. Preliminary results on PSR J1930+1852 were also presented at the conference. This small dataset (5 hours) consisted of observations made mainly under extremely bright moonlight, and is currently being reanalysed with an analysis and Monte Carlo chain more suitable for these conditions. Additionally, this object is in the galactic plane and so does not fullfill the criteria of the present study. 
\begin{table}
\begin{tabular}{lccccccc}
\hline
   \tablehead{1}{r}{b}{PSR name}
  & \tablehead{1}{r}{b}{log$_{10}\dot{E}/d^{2}$ [erg/s/kpc$^2$]}
  & \tablehead{1}{r}{b}{T [hrs]}
  & \tablehead{1}{r}{b}{Z [deg]}
  & \tablehead{1}{r}{b}{S [$\sigma$]}   
  & \tablehead{1}{r}{b}{F($>E_{th}$) [f.u.]}   
  & \tablehead{1}{r}{b}{F($>E_{th}$) [\% Crab]}   \\
\hline
J0205+6449 & 36.4 & 12.8 & 35.2  & 1.1 & 2.9 & 2.3  \\
J0631+1036 & 35.2 & 13.0 & 24.4 & 0.3  & 1.6 & 1.3 \\
J0633+1746 & 36.1 & 14.5 & 17.9 & 1.1  & 2.0 & 1.6 \\
B0656+14 & 35.6 & 9.4 & 22.4 & -1.8  & 0.3 & 0.2 \\
J1740+1000 & 35.1 & 10.5 & 24.6 & 0.2  & 1.2 & 1.0 \\
\hline
\end{tabular}
\caption{Summary of the VERITAS observations (livetime after dead time correction, mean zenith angle) and results (pre-trials significance, 99\% C.L. upper limits on the integral flux emission above 300 GeV in f.u.=10$^ {-12}$ cm$^{-2}$s$^{-1}$ and Crab units) for the selection of pulsars
\label{tab:a}}
\end{table}
\section{Observations, analysis and results}
VERITAS is a system of four, 12 m diameter Imaging Atmospheric Cherenkov Telescopes (IACTs) dedicated to the observation of TeV gamma rays. Fully operational since April 2007, VERITAS is located at a a latitude 32$^{\circ}$N and longitude 111$^{\circ}$W, providing best visibility to the Northern sky. The array can detect $\gamma$-rays with high sensitivity (5$\sigma$ detection in 50 h for 1\% Crab flux) and an excellent angular (0.1 deg) and energy (15-20\%) resolution over a $\sim4^{\circ}$ field of view and an energy range of 100 GeV to 30 TeV. This performance makes VERITAS suitable for spectral and morphology studies of extended sources and/or of sources with uncertain positions, typical of PWNe. 
\par
In September 2007 the full VERITAS array started the program to look for VHE $\gamma$-ray counterparts of the most energetic pulsars in the Northern sky outside of the galactic plane region covered by the VERITAS sky survey. The observations presented here are summarized in Table 1. The live time of the quality selected and dead time corrected dataset for each object ranges from 9 to 15 hours at small and medium zenith angles. All data were taken in ``wobble'' mode~\cite{Daum97} with the source candidate offset by 0.5$^{\circ}$ from the centre of the field of view. 
\par
The standard scheme for the reconstruction of events used in VERITAS was applied to the data\cite{Acciari08a}. This reconstruction is based on image parameters\cite{Hillas85}. Selection of the events optimized for a 1\% Crab flux were used to suppress the otherwise overwhelming hadronic background. Events with four telescope images are those giving the maximum signal-to-noise ratio and a better angular resolution at expenses of a slightly increase of energy threshold. The background estimation for each position in the two-dimensional sky maps is computed from a ring with a radius of 0.5$^{\circ}$. Since the emission region may be extended, the search for a $\gamma$-ray signal is performed with  two fixed integration radii ($\theta_{int}=0.13^ {\circ}$ and $0.2^ {\circ}$). These trials need to be taken into account in assessing the true chance probability of any observed excess. 
\begin{figure}[t!]
\includegraphics[width=6cm]{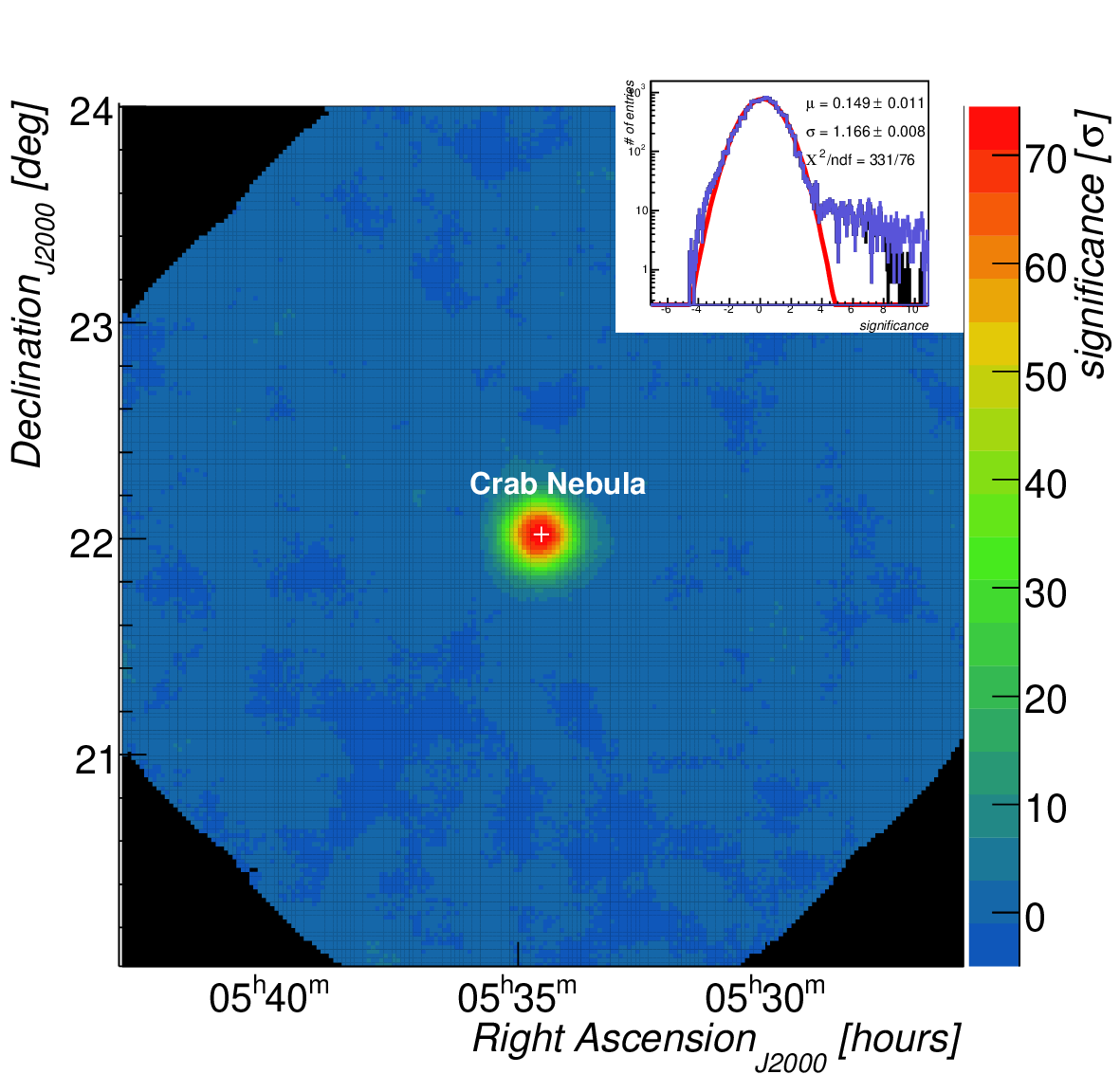}
\caption{\label{CrabPlot} Calculated significance across from the FOV containing the Crab Nebula as observed with VERITAS. The significance distribution of the various bins in the vicinity is also plotted showing}
\end{figure}
\begin{figure}[t!]
\begin{minipage}[b]{0.8\linewidth}
\includegraphics[width=6cm]{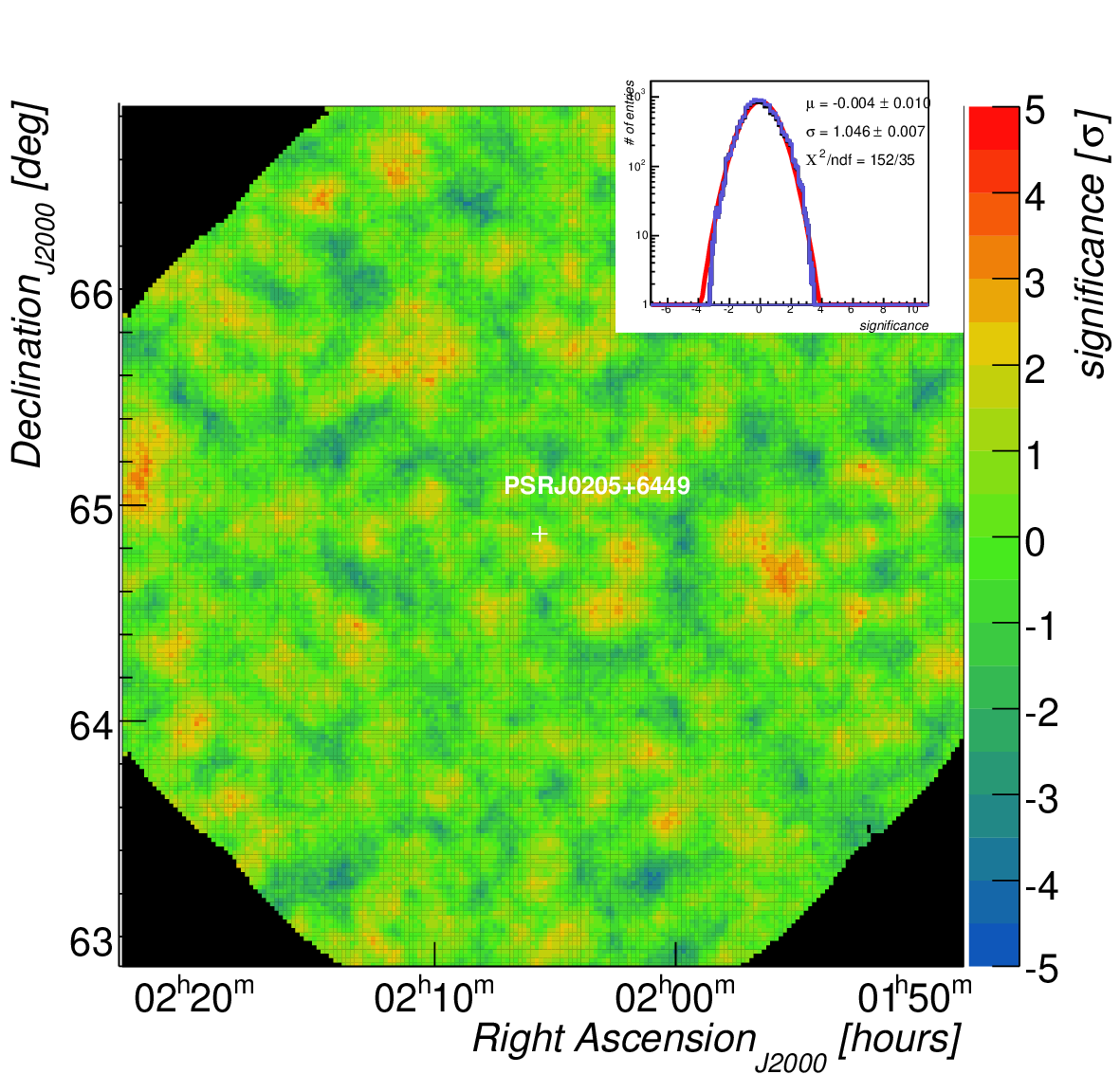}
\includegraphics[width=6cm]{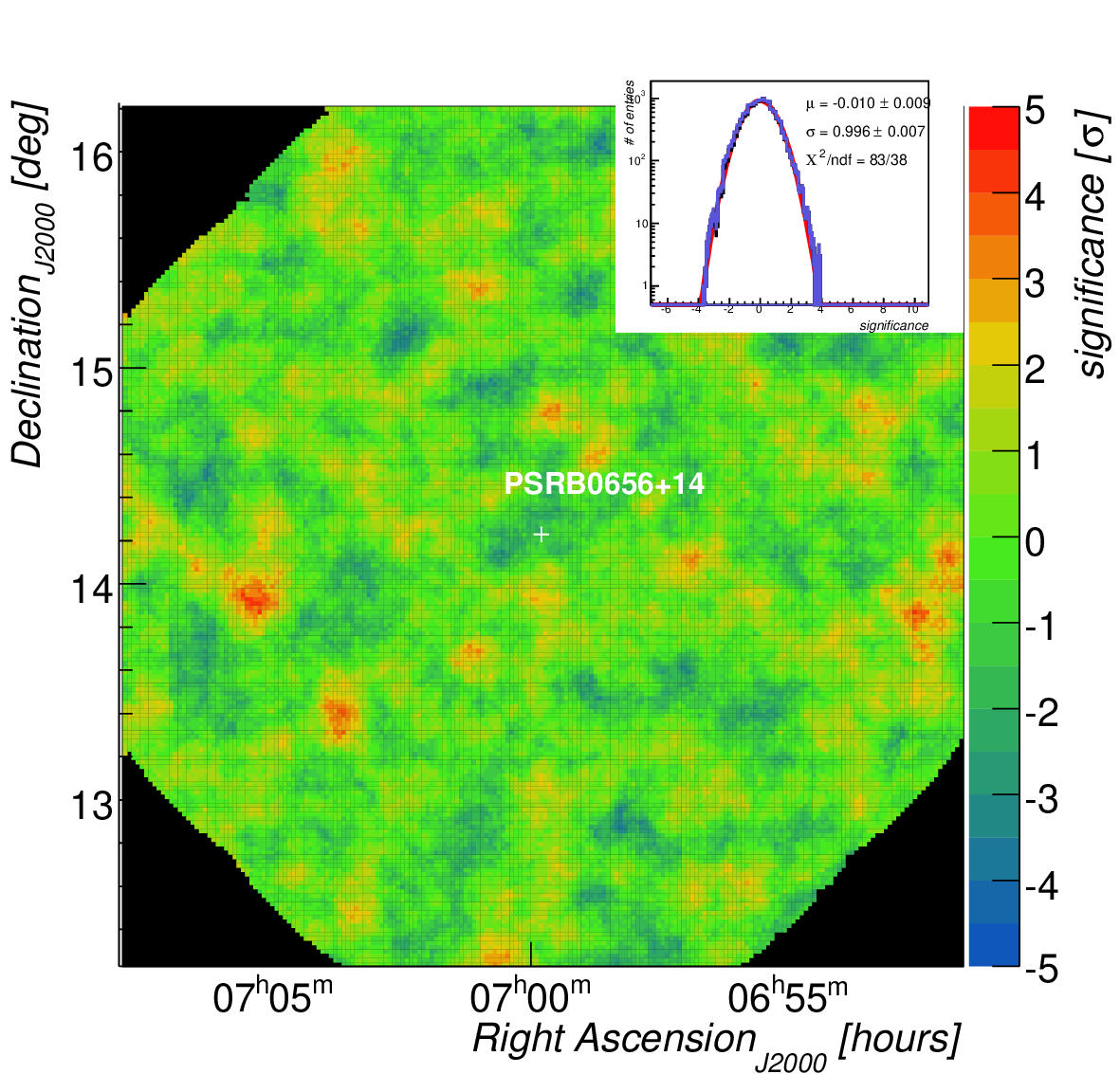}
\end{minipage}
\begin{minipage}[b]{0.8\linewidth}
\includegraphics[width=6cm]{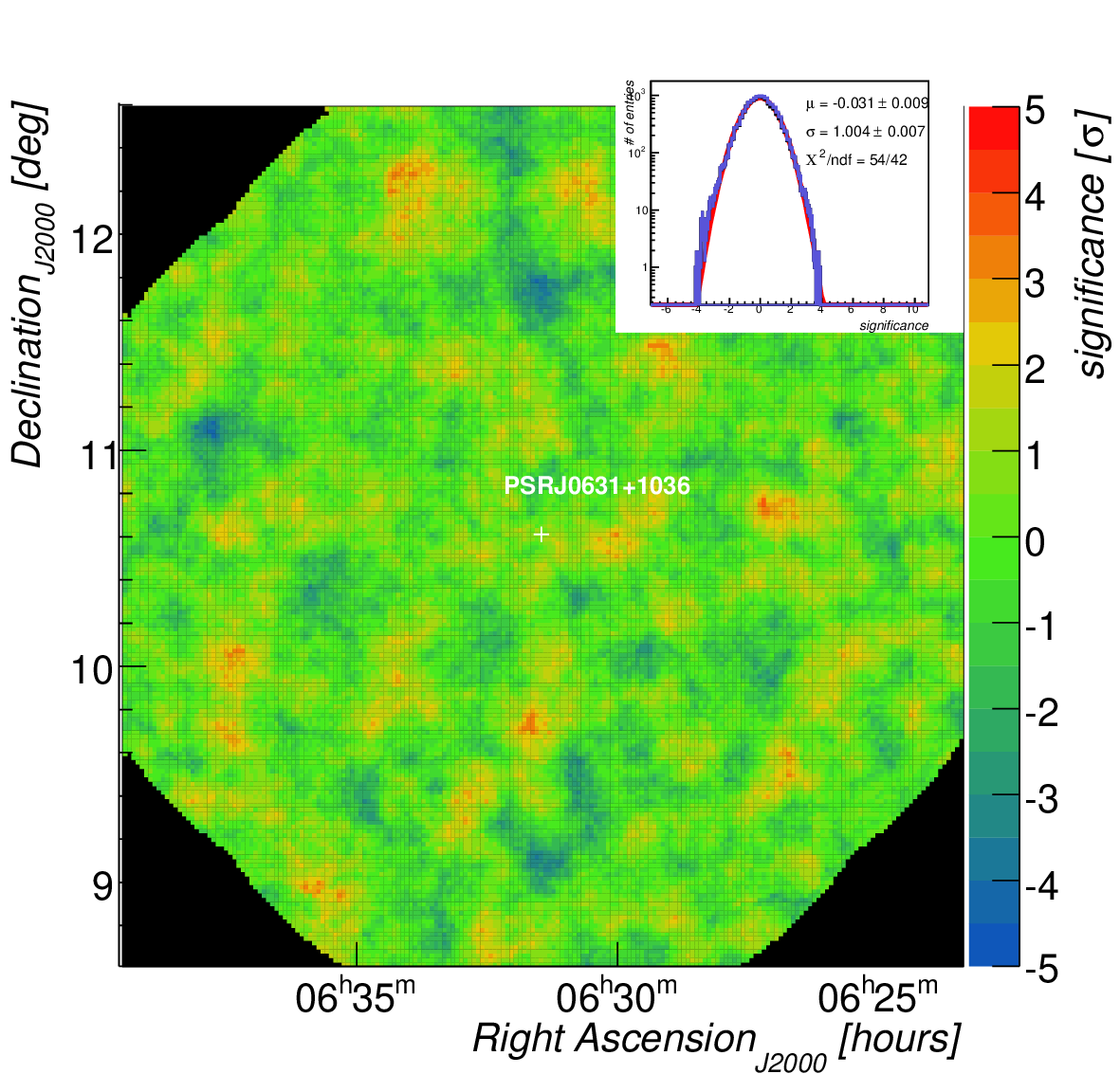}
\includegraphics[width=6cm]{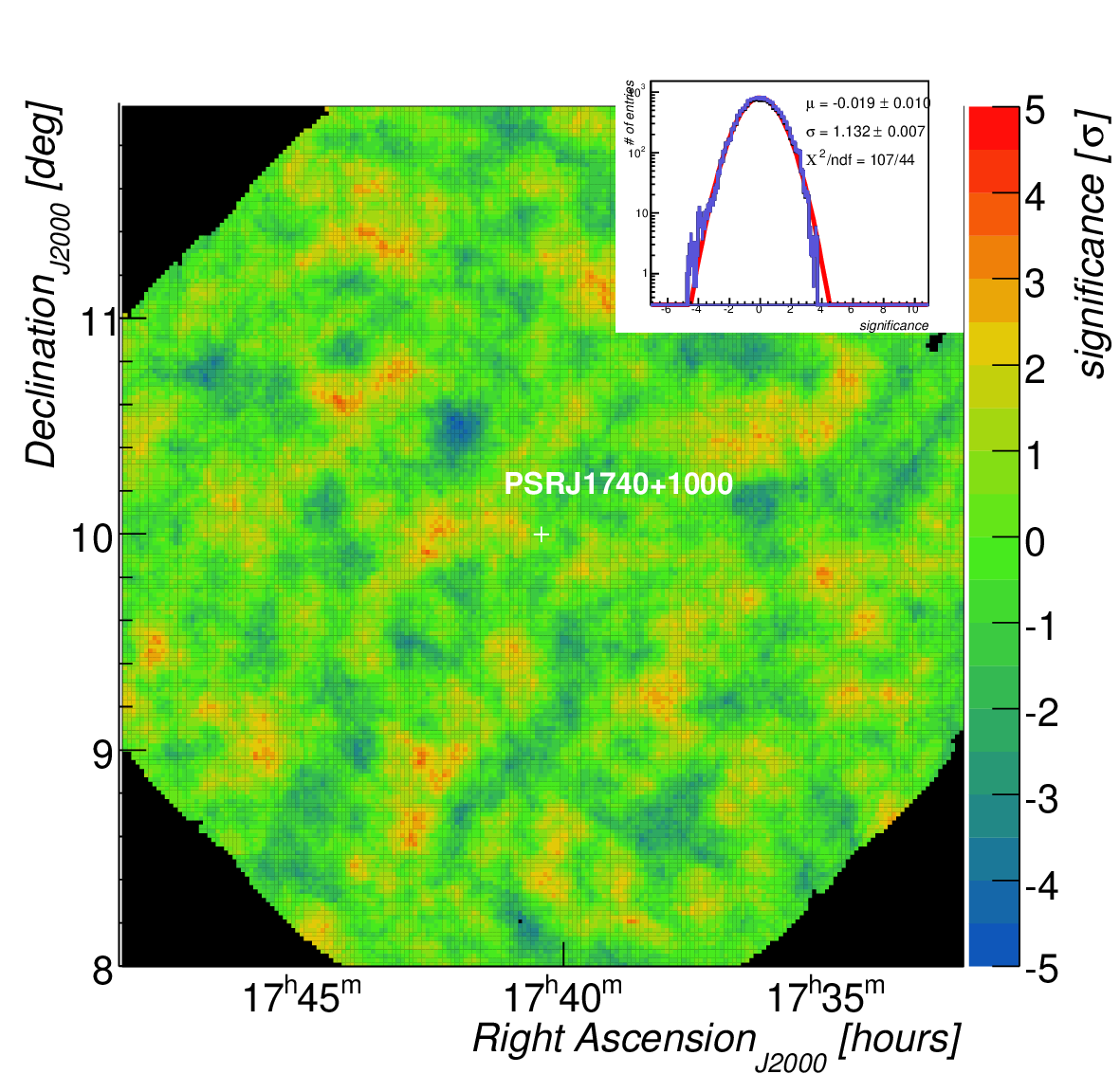}
\end{minipage}
\caption{\label{plots} Calculated significance across from the FOV containing each of the studied pulsars observed with VERITAS (marked with the white crosses). For each map the distribution of significances in the vicinity of the pulsar is also indicated.}
\end{figure}
\par
Fig.~\ref{plots} shows the smoothed significance maps of the studied pulsars and their surroundings after sampling with a $\theta_{int}=0.13^{\circ}$, corresponding to the point source search. None of the maps show a significant VHE $\gamma$-ray signal; the distribution of significances over all bins is compatible with randomly distributed data ($\mu\sim0$,$\sigma\sim1$). Similar results are found for the wider integration radius. For comparison the significance map of Crab Nebula observations with VERITAS using the same observation mode and analysis is also shown in Fig.~\ref{CrabPlot}. A significance of 75$\sigma$ in $\sim7.5$ h of VERITAS observations at small zenith angles is measured.
\par
Results from this search are summarized in in Table 1. The significances of the VHE $\gamma$-ray signal at the pulsar position are given for an assumed point source. Negative significances result purely from fluctuations in the cosmic-ray background and should not be interpreted as a genuine deficit in the photon flux. Upper limits on the steady flux emission above 300 GeV  at the 99\% confidence level have been derived using the prescription of Rolke~\cite{Rolke} for each of the pulsars. These upper limits have been calculated assuming a point source and assuming a power-law spectrum with photon index of $\Gamma=2.5$ ($\sim E^{-\Gamma}$) for the calculation of the effective collection area. The limits range from 0.2\%(for negative significances) to 2.3\% Crab. We note that care must be taken when interpreting these limits as most of the detected middle age pulsars (all the cases here except J0205+6449) drive extended TeV PWNe. Future work will provide more detailed extended source results.

\section{Conclusions}
The search for  VHE $\gamma$-ray counterparts of 5 energetic and/or close pulsars at Northern latitudes has shown no detection by VERITAS. Upper limits have been derived at the 99\% confidence level assuming a point source. Taking into account the nature of the objects being searched (moderate extended $\sim$0.2 deg, TeV nebulae close to pulsars) and the good angular resolution and sensitivity of VERITAS observations compared to other Northern instruments, this represents the most comprehensive search for VHE $\gamma$-rays to date around those objects of the high Galactic latitudes.
\par
The absence of any detection in these objects could be explained in a number of ways:
\begin{itemize}
\item The exposure on those objects could be not enough and some of the most promising objects could be re-observed in the future
\item Given the middle age of the observed pulsars (except J0205+6449), can happen that the PWN is not inside the SNR and therefore it is confined by the ambient ISM. A low density of the ISM could difficult the confinement of the wind particles to form the nebulae in these objects. A study of the ISM around these objects will help to determine if they reside in a lower density region of the ISM compared to some of the lower Galactic latitude detections.
\item  Another reason could be the fact that some of the HESS detections have been produced close to the center of the galactic plance where the IC component due to the star dust can be more important.
\end{itemize}  
\par
Observations and analysis of more energetic pulsars with VERITAS is still ongoing and more extensive results will be presented in the future.


\begin{theacknowledgments}
This research is supported by grants from the U.S. Department of Energy, the U.S. National Science Foundation and the Smithsonian Institution, by NSERC in Canada, by PPARC in the U.K. and by Science Foundation Ireland. 
\end{theacknowledgments}



\bibliographystyle{aipproc}   

\end{document}